\documentclass[prb,twocolumn,showpacs]{revtex4}
\usepackage{amsmath,amsthm,amssymb,amscd,float}
\usepackage{psfrag,graphicx}
\newcommand{\al}{\alpha}
\newcommand{\be}{\beta}
\newcommand{\de}{\delta}

\newcommand{\ga}{\gamma}

\newcommand{\si}{\sigma}

\newcommand{\De}{\Delta}

\newcommand{\La}{\Lambda}

\newcommand{\bk}{\mathbf{k}}

\newcommand{\bp}{\mathbf{p}}
\newcommand{\bs}{\mathbf{s}}
\newcommand{\bx}{\mathbf{x}}

\newcommand{\tK}{\widetilde{K}}
\newcommand{\tP}{\widetilde{P}}

\newcommand{\tk}{\tilde{k}}

\newcommand{\ssH}{\mathsf{H}}

\newcommand{\hp}{\hat{p}}

\newcommand{\NN}{{\mathbb N}}
\newcommand{\RR}{{\mathbb R}}

\newcommand{\cP}{{\mathcal P}}

\newcommand{\cY}{{\mathcal Y}}

\def\BE{\,\overline{\!E}{}}
\def\BH{\,\overline{\!H}{}}

\newcommand{\pa}{\partial}

\def\ket#1{|#1\rangle}

\let\ni\noindent
\newcommand{\ms}{\mspace{1mu}}
\newcommand{\erf}{\operatorname{erf}}

\newcommand{\tr}{\operatorname{tr}}

\newcommand{\iu}{\mathrm{i}}
\newcommand{\e}{\mathrm{e}}
\newcommand{\Or}{\mathrm{O}}
\newcommand{\diff}{\mathrm{d}}
\newcounter{ex}
\def\cond{\stepcounter{ex}\hskip-.8cm
\makebox[.6cm][r]{\roman{ex})}\hskip.2cm}
\begin{document}
\title{Global properties of the spectrum of the Haldane--Shastry spin chain}
\author{Federico \surname{Finkel}}%
\email{ffinkel@fis.ucm.es}
\author{Artemio \surname{Gonz{\'a}lez-L{\'o}pez}}
\email[Corresponding author. Electronic address: ]{artemio@fis.ucm.es}
\affiliation{Departamento de F{\'\i}sica Te{\'o}rica II,
Universidad Complutense, 28040 Madrid, Spain}
\date{May 23, 2005; revised September 1, 2005}
\begin{abstract}
  We derive an exact expression for the partition function of the
  $\mathrm{su}(m)$ Haldane--Shastry spin chain, which we use to study the
  density of levels and the distribution of the spacing between consecutive
  levels. Our computations show that when the number of sites $N$ is large
  enough the level density is Gaussian to a very high degree of approximation.
  More surprisingly, we also find that the nearest-neighbor spacing
  distribution is not Poissonian, so that this model departs from the typical
  behavior for an integrable system. We show that the cumulative spacing
  distribution of the model can be well approximated by a simple functional
  law involving only three parameters.
\end{abstract}
\maketitle
%
\section{Introduction}

The Haldane--Shastry (HS) spin chain describes $N$ spins equally
spaced on a circle with an interaction inversely proportional to
the square of their chord distance~\cite{Ha88,Sh88}. The original
motivation for studying this model is the fact that it possesses
an exact Jastrow-product ground state, which coincides with the
$U\to\infty$ limit of Gutzwiller's variational wave function for
the Hubbard model~\cite{Gu63,GV87,GJR87}, and also with the
one-dimensional version of the resonating valence bond state
introduced by Anderson~\cite{ABZH87}. Since its very introduction,
the HS spin chain has been extensively studied as a completely
integrable model~\cite{FM93} solvable by the asymptotic Bethe
ansatz~\cite{Ha91,Ka92,HH93}, whose spinon excitations provide a
simple example of a system obeying fractional
statistics~\cite{Ha91b}.

The energy spectrum of the HS Hamiltonian with spin $1/2$ was
partially computed in the original papers of Haldane and Shastry.
In a subsequent publication~\cite{HHTBP92}, Haldane et
al.~empirically found a complete description of the spectrum for
arbitrary spin, and explained its highly degenerate character by
the symmetry of the model under the Yangian algebra
$\cY(\mathrm{sl}_m)$. These results were rigorously established in
Ref.~\cite{BGHP93} by explicitly constructing a transfer matrix in
terms of the Dunkl operators~\cite{Du89,Po92} of the trigonometric
Sutherland dynamical model~\cite{Su71,Su72}. In this approach, the
spectrum is obtained by considering all possible \emph{motifs}
$\de\equiv(0\ms\de_1\dots\de_{N-1}\ms 0)$, where each $\de_j$ is
either $0$ or $1$ and the maximum number of consecutive $1$'s is
$m-1$. Indeed, the energy associated with a motif $\de$ is given
by the compact formula
\begin{equation}\label{Ep}
E_{\mathrm{HS}}(\de)=\sum_{j=1}^{N-1}\de_j\ms j(j-N)\,.
\end{equation}
The degeneracy of a level $E_\mathrm{HS}$ is obtained by summing the
degeneracies corresponding to all the motifs $\de$ such that
$E_\mathrm{HS}(\de)=E_\mathrm{HS}$. Although there is a well-defined algorithm for
computing the degeneracy of each motif, in practice the
computation becomes quite involved except for $m=2$. It is therefore
difficult to derive in this way an exact expression for the
partition function valid for arbitrary values of $N$ and $m$. Perhaps as a
consequence of this fact, little attention has been paid in the
literature to the global properties of the spectrum of the HS
chain.

Some authors~\cite{HB00,HW03} have suggested that the main
obstacle in computing the partition function of the HS chain in
closed form is the fact that the dispersion relation~\eqref{Ep} is
nonlinear in $j$, in contrast with the Polychronakos rational
chain~\cite{Fr93,Po93}. In a recent paper~\cite{EFGR05}, however,
the partition function of the trigonometric HS spin chain of
$BC_N$ type has been exactly computed applying what is known as Polychronakos's
\emph{freezing trick}~\cite{Po94}, notwithstanding the fact that
these chains have a nonlinear dispersion relation similar
to~\eqref{Ep}. In fact, we shall prove in what follows that the
partition function of the chain~\eqref{HS} can also be computed
using the freezing trick. {}From the partition function it is
straightforward to generate the spectrum of the HS chain for a
wide range of values of $N$ and $m$, and thus study global
properties thereof such as the level density or the distribution
of the spacing between consecutive levels.

\section{Partition function}

For convenience, we shall take the Hamiltonian of the (antiferromagnetic)
Haldane--Shastry spin chain as
\begin{equation}\label{HS}
H=\frac12\sum_{i<j}\sin(\xi_i-\xi_j)^{-2}(1+S_{ij})\,,
\end{equation}
where $\xi_i=i\pi/N$ and $S_{ij}$ is the spin permutation operator
of particles $i$ and $j$. Here and throughout the paper all sums and products
run from $1$ to $N$ unless otherwise specified. The Hamiltonian of the original HS spin
chain is given by $H_{\mathrm{HS}}=H-E_{\mathrm{max}}$, where
\begin{equation}\label{Emaxdef}
E_{\mathrm{max}}\equiv\sum_{i<j}\sin(\xi_i-\xi_j)^{-2}
\end{equation}
is the highest energy of $H$. In order to apply the freezing
trick, we need to introduce the Sutherland spin model
\begin{equation}
\label{Hstar}
  H^*=-\sum_i \pa_{x_i}^2+ a\,\sum_{i\neq j}\sin(x_i-x_j)^{-2}\,(a+S_{ij})\,,
\end{equation}
and its scalar version
\[
H_0=-\sum_i \pa_{x_i}^2+ a(a-1)\,\sum_{i\neq j}\sin(x_i-x_j)^{-2}\,.
\]
We thus have
\begin{equation}
\label{h}
  H^*=H_0+4a\ssH\,,
\end{equation}
where $\ssH$ is obtained from $H$ by the replacement $\xi_i\to x_i$.
The freezing trick is based on the fact that for $a\to\infty$ the particles
``freeze'' at the equilibrium positions of the scalar part of the potential in $H^*$,
which are simply the lattice points of the chain~\eqref{HS}. In this limit,
the spin degrees of freedom decouple from the dynamical ones, so that by Eq.~\eqref{h} the
energies of the dynamical spin model are approximately given by~\cite{EFGR05}
\begin{equation}\label{Eij}
E^*_{ij}\simeq E_{0,i}+4a\ms E_j\,,
\end{equation}
where $E_{0,i}$ and $E_j$ are \emph{any} two levels of $H_0$ and $H$. Hence
the partition functions $Z$, $Z^*$, and $Z_0$ of $H$, $H^*$, and $H_0$,
respectively, satisfy the approximate equality
\[
Z^*(T)\simeq Z_0(T)Z\big({\textstyle\frac T{4a}}\big)\,,
\qquad a\gg1\,.
\]
The latter equation leads to the \emph{exact} formula
\begin{equation}
\label{Z}
Z(T)=\lim_{a\to\infty}\frac{Z^*(4aT)}{Z_0(4aT)}\,,
\end{equation}
which we will use to compute the partition function of the chain~\eqref{HS} in closed form.

In order to evaluate the RHS of~\eqref{Z}, we need to compute the
spectra of $H^*$ and of its scalar limit $H_0$. These spectra can
be obtained in a unified way by considering the scalar
differential-difference operator
\begin{equation}\label{BH}
\BH=-\sum_i \pa_{x_i}^2+ a\,\sum_{i\neq j}\sin(x_i-x_j)^{-2}\,(a-P_{ij})\,,
\end{equation}
where $P_{ij}$ permutes the coordinates $i$ and $j$. The
operator $\BH$ is represented by an upper triangular matrix in a
(non-orthonormal) basis whose elements are of the form
\begin{equation}\label{phis}
\phi_{\bp}(\bx)=\e^{2\iu\bp\cdot\bx}\prod_{i<j}\sin^a(x_i-x_j)\,,
\end{equation}
where the vector $\bp=(p_1,\dots,p_N)\in\RR^N$ is such that the
differences $p_i-p_{i+1}$, $1\leq i\leq N-1$, are integers. The
basis elements~\eqref{phis} should be ordered in a suitable way
that we shall now describe. We shall say that a vector
$\hat\bp=(\hp_1,\dots,\hp_N)$ is \emph{nonincreasing} if
$\hp_{i+1}\leq\hp_i$ for $i=1,\dots,N-1$. Given two nonincreasing
vectors $\hat\bp$ and $\hat\bp'$, we shall write
$\hat\bp\prec\hat\bp'$ if
$\hp_1-\hp_1'=\cdots=\hp_{i-1}-\hp_{i-1}'=0$ and $\hp_i<\hp_i'$.
Finally, we say that the basis element $\phi_{\bp}$ precedes
$\phi_{\bp'}$ if $\hat\bp\prec\hat\bp'$, where $\hat\bp$ and
$\hat\bp'$ are the unique nonincreasing vectors obtained from
$\bp$ and $\bp'$ by reordering their components. It can then be shown
that the matrix of $\BH$ in the basis $\{\phi_\bp\}$ with the
order just defined is indeed upper triangular, with diagonal
elements~\cite{BGHP93,Ba96}
\begin{equation}\label{BE}
\BE(\bp)=\sum_i \big(2\hp_i+a(N+1-2i)\big)^2\,.
\end{equation}

We shall now see how the spectrum of $H^*$ follows easily from
that of $\BH$. To this end, let us introduce the total
antisymmetrizer $\La$ with respect to simultaneous permutations of
the spatial and spin coordinates. We can construct a
(non-orthonormal) basis of the Hilbert space of the Hamiltonian
$H^*$ with states of the form
\begin{equation}\label{psis}
\psi_{\bp,\bs}(\bx)=\La\big(\phi_\bp(\bx)\ket\bs\big)\,,
\end{equation}
where $\ket\bs\equiv\ket{s_1,\dots,s_N}$ is an element of the spin
basis and the vector $\bp$ satisfies the following
conditions:\smallskip

{\leftskip.8cm\parindent=0pt%
\cond The differences $n_i\equiv p_i-p_{i+1}$, $1\leq i\leq N-1$,
are nonnegative integers.

\cond At most $m$ components of $\bp$ can be equal.\par

\cond The total momentum vanishes, i.e., $\sum_i p_i=0$.\smallskip

} \ni The first two conditions are a direct consequence of the
antisymmetric nature of the states~\eqref{psis}. The last
condition reflects the fact that, since $H^*$ is translationally
invariant, we can work in the center of mass frame. The basic
states $\psi_{\bp,\bs}$ should be ordered in such a way that
$\psi_{\bp,\bs}$ precedes $\psi_{\bp'\!,\bs'}$ if $\bp\prec\bp'$
(note that the vectors $\bp$ and $\bp'$ are nonincreasing by
condition i)).

{}From the elementary relations $P_{ij}\La=-S_{ij}\La$ and the
fact that $\BH$ clearly commutes with $\La$, it follows that
\begin{align*}
H^*\psi_{\bp,\bs}&=\BH\psi_{\bp,\bs}
=\La\big((\BH\phi_\bp)\ket\bs\big)\\[1mm]
&=\La\Big(\BE(\bp)\phi_\bp\ket\bs+\sum_{\bp'\prec\bp}c_{\bp\bp'}\phi_{\bp'}\ket\bs\Big)\\
&=\BE(\bp)\psi_{\bp,\bs}+\sum_{\bp'\prec\bp}c_{\bp\bp'}\psi_{\bp'\!,\bs}\,.
\end{align*}
Hence the Hamiltonian $H^*$ of the Sutherland spin model is upper
triangular in the basis $\{\psi_{\bp,\bs}\}$, with diagonal
elements
\begin{equation}\label{E*}
E^*(\bp,\bs)=\sum_i \big(2p_i+a(N+1-2i)\big)^2\,,
\end{equation}
where $\bp$ satisfies conditions i)--iii) above.

The spectrum of $H_0$ can be derived by a similar argument, noting
that $H_0=\BH$ on scalar symmetric states of the form
$\psi_\bp=\La_{\mathrm s}\phi_\bp$, where $\La_{\mathrm s}$ is the
symmetrizer with respect to the spatial coordinates and $\bp$
satisfies only conditions i) and iii) above. Hence~\cite{Su72} the
eigenvalues $E_0(\bp)$ of $H_0$ are also given by the RHS
of~\eqref{E*}, where now $\bp$ is not restricted by condition ii).

{}From the above results it is easy to compute the partition
functions $Z_0(4aT)$ and $Z^*(4aT)$ in the limit $a\to\infty$. For
the computation of $Z_0(4aT)$, we start by expanding the eigenvalues of $H_0$
in powers of $a$ as
\begin{equation}\label{E0a}
E_0(\bp)=a^2E^0+4a\sum_i(N+1-2i)p_i+\Or(1)\,,
\end{equation}
where
\[
E^0=\sum_i(N+1-2i)^2=\frac13\,N(N^2-1).
\]
Since $E^0$ does not depend on $\bp$, and therefore contributes the same
overall constant factor to both $Z_0$ and $Z^*$, we shall henceforth drop the
first term in Eq.~\eqref{E0a}. With this convention, for $a\gg1$
the denominator in Eq.~\eqref{Z} is given by
\[
Z_0(4aT)\simeq \sum_\bp q^{\sum_i p_i(N+1-2i)}\,,
\]
where $q=\e^{-1/(k_{\mathrm{B}}T)}$ and the outer sum runs over
all vectors $\bp$ satisfying conditions i) and iii) above. Setting
$n_N\equiv p_N$ we have
\[
\sum_i p_i(N+1-2i)=\sum_{j\geq i}n_j(N+1-2i)=\sum_{j=1}^{N-1}j(N-j)n_j\,.
\]
Taking into account that $n_N$ is determined by the remaining
$n_i$'s by condition iii), we finally obtain
\begin{multline}\label{Z0}
Z_0(4aT)\simeq \sum_{n_1,\dots,n_{N-1}\geq0}
\,\prod_{j=1}^{N-1}q^{j(N-j)n_j}\\
=\prod_{j=1}^{N-1}\Big(1-q^{j(N-j)}\Big)^{-1}\,.
\end{multline}

In order to compute the partition function $Z^*(4aT)$ for
$a\gg 1$, it is convenient to represent the vector $\bp$ labeling the
energies~\eqref{E*} of $H^*$ as
\begin{equation}
\label{nm} \bp =
\big(\overbrace{\vphantom{1}\rho_1,\dots,\rho_1}^{k_1},\dots,
\overbrace{\vphantom{1}\rho_r,\dots,\rho_r}^{k_r}\big).
\end{equation}
Note that $\sum\limits_{i=1}^r k_i=N$, so that
$\bk=(k_1,\dots,k_r)$ belongs to the set $\cP_N$ of
partitions of $N$ (taking order into account). Calling
\begin{equation}\label{Ki}
K_i=\sum\limits_{j=1}^i k_j,
\end{equation}
and dropping again the term $a^2E^0$, in the large $a$ limit Eq.~\eqref{E*} becomes
\[
E^*(\bp,\bs)\simeq 4a\sum_{i=1}^r
\rho_i\sum_{j=K_{i-1}+1}^{K_i}(N+1-2j)=4a\sum_{i=1}^r \rho_il_i\,,
\]
where
\begin{equation}\label{li}
l_i=k_i(N-2K_i+k_i).
\end{equation}
Since $E^*(\bp,\bs)$ does not depend on the spin coordinates~$\bs$,
the degeneracy associated with this eigenvalue
is given by
\[
d(\bk)=\prod\limits_{i=1}^r\binom m{k_i},
\]
so that $d(\bk)=0$ if $k_i>m$ for some $i$, in accordance with condition ii).
Hence
\begin{equation}\label{Z*1}
Z^*(4aT)\simeq \sum_{\bk\in\cP_N}d(\bk)\!
\sum_{\substack{\rho_1>\cdots>\rho_r\\[2pt]k_1\rho_1+\cdots+k_r\rho_r=0}}\!
q^{\,\sum\limits_{i=1}^r\rho_il_i}.
\end{equation}
Calling $\nu_i=\rho_{i}-\rho_{i+1}\in\NN$, $i=1,\dots,r-1$, and $\nu_r=\rho_r$,
we have
\begin{equation}\label{nuili}
\sum_{i=1}^r\rho_i l_i=\sum_{1\leq i\leq j\leq r}l_i\nu_j=\sum_{j=1}^r\nu_jN_j\,,
\end{equation}
where
\[
N_j=\sum_{i=1}^jl_i=K_j(N-K_j)\,,
\]
by Eq.~\eqref{li}. Note, in particular, that the numbers $N_j$ depend on $\bk$
through the partial sums~\eqref{Ki}.
Substituting~\eqref{nuili} into~\eqref{Z*1}, and taking into account
that $K_r=N$ implies $N_r=0$, we obtain
\begin{multline}\label{Z*}
Z^*(4aT)\simeq \sum_{\bk\in\cP_N}d(\bk)
\sum_{\nu_1,\dots,\nu_{r-1}>0}\prod_{j=1}^{r-1}
q^{N_j\nu_j}\\
=\sum_{\bk\in\cP_N}d(\bk)\prod_{j=1}^{r-1}\frac{q^{N_j}}{1-q^{N_j}}\,.
\end{multline}
Combining Eqs.~\eqref{Z0} and~\eqref{Z*}, the partition function $Z$
can be expressed in closed form as
\begin{equation}\label{Z1}
Z(T)=\prod_{j=1}^{N-1}\Big(1-q^{j(N-j)}\Big)
\sum_{\bk\in\cP_N}d(\bk)
\prod_{i=1}^{r-1}\frac{q^{N_i}}{1-q^{N_i}}\,.
\end{equation}
Note that, by definition, the partial sums $K_i$ are natural numbers satisfying
$1\leq K_1<\cdots<K_{r-1}\leq N-1$. Denoting by $K'_1<\dots<K'_{N-r}$
the elements of the set
\[
\{1,\dots,N-1\}-\{K_1,\dots,K_{r-1}\}\,,
\]
and setting
\[
N'_i=K'_i(N-K'_i)\,,
\]
we have
\[
\prod_{j=1}^{N-1}\Big(1-q^{j(N-j)}\Big)
=\prod_{i=1}^{r-1}\big(1-q^{N_i}\big)\prod_{i=1}^{N-r}\big(1-q^{N'_i}\big)\,.
\]
This identity and Eq.~\eqref{Z1} yield the following remarkable formula
for the partition function of the spin chain~\eqref{HS}:
\begin{equation}\label{Zfinal}
Z(T)=\sum_{\bk\in\cP_N}\prod_{i=1}^r\binom{m}{k_i}\,
q^{\,\sum\limits_{i=1}^{r-1}N_i}
\prod_{i=1}^{N-r}\big(1-q^{N'_i}\big).
\end{equation}

{}From the previous formula it follows that the energy levels of $H$ are of the
form
\begin{equation}\label{Edep}
E(\de')=\sum_{j=1}^{N-1}\de'_{\!j}\, j(N-j)\,,
\end{equation}
where $\de'_{\!j}=1$ if $j$ is one of the partial sums $\tK_i$
corresponding to a partition $(\tk_1,\dots,\tk_r)\in\cP_N$ with
$\tk_l\leq m$ for all $l$ (by condition ii)), and $\de'_{\!j}=0$ otherwise.
In order to relate Eq.~\eqref{Edep} with the known expression~\eqref{Ep}
for the energies of the original HS Hamiltonian, we
need to evaluate the maximum energy $E_\mathrm{max}$. {}From Eq.~\eqref{Emaxdef} we have
\begin{align*}
E_\mathrm{max}&=\sum_{j=1}^{N-1}(N-j)\csc^2\Big(\frac{j\pi}N\Big)\\
&=\sum_{j=1}^{N-1}j\,\csc^2\bigg(\frac{(N-j)\pi}N\bigg)
=\sum_{j=1}^{N-1}j\,\csc^2\Big(\frac{j\pi}N\Big).
\end{align*}
Hence
\begin{equation}\label{Emax}
E_\mathrm{max}=\frac N2\sum_{j=1}^{N-1}\csc^2\Big(\frac{j\pi}N\Big)=\frac N6(N^2-1)\,,
\end{equation}
where the last sum is evaluated in Ref.~\cite{CP78}. Since the RHS of~\eqref{Emax}
coincides with the sum $\sum_{j=1}^{N-1}j(N-j)$,
Eq.~\eqref{Edep} implies Eq.~\eqref{Ep} with $\de_j=1-\de'_{\!j}$. In particular,
from the latter relation between $\de$ and $\de'$ it follows that $\de$
is a motif with no more than $m-1$ consecutive $1$'s.

\section{Level density and spacings distribution}

\begin{figure}[h]
\psfrag{F}[Bc][Bc][1][0]{\begin{footnotesize}$F(E),\,G(E)$\end{footnotesize}}
\psfrag{E}{\begin{footnotesize}$E$\end{footnotesize}}
\includegraphics[width=8cm]{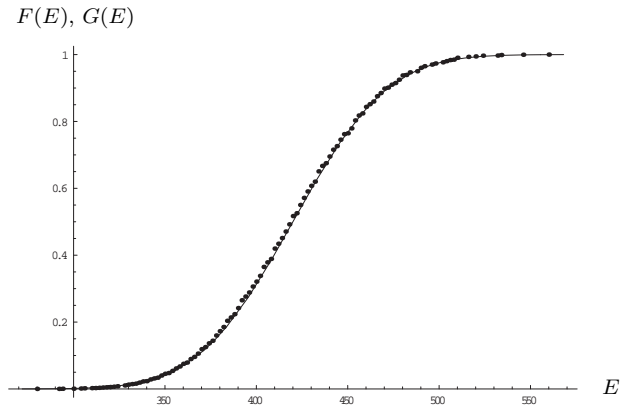}
\caption{Cumulative distribution functions $F(E)$ (at its discontinuity
points) and $G(E)$ (continuous line) for $N=15$ and $m=2$.\label{levden}}
\end{figure}

The RHS of Eq.~\eqref{Zfinal} is a polynomial in $q$ whose evaluation with a
symbolic algebra package is straightforward once $N$ and $m$ are fixed. In
this way we have been able to compute the spectrum of the chain~\eqref{HS} for
relatively large values of $N$ and $m$, for which the usual motif approach
becomes inefficient due to the difficulty of computing the degeneracies.
{}From the analysis of the spectral data thus obtained one can infer several
global properties of the spectrum that we shall now discuss. In the first
place, it is apparent that for $N\gg 1$ the level density is Gaussian to a very
high degree of accuracy, as in the HS spin chain of $\mathrm{BC}_N$ type
studied in Ref.~\cite{EFGR05}. In other words, for large $N$ the cumulative
level density
\[
F(E)=m^{-N}\sum\limits_{i;E_i\leq E}d_i
\]
is approximately given by
\[
G(E)=\frac12\bigg[1+\erf\bigg(\frac{E-\mu}{\sqrt 2\si}\,\bigg)\bigg],
\]
where $d_i$ is the degeneracy of the energy $E_i$, and $\mu$ and $\si$ are
respectively the mean and the standard deviation of the energy. This can
already be seen, for instance, in the case $N=15$ and $m=2$ presented in
Fig.~\ref{levden}. The agreement between $F$ and $G$ rapidly improves as $N$
and/or $m$ grow, e.g., for $m=2$ the mean square error decreases from $5.2\times
10^{-5}$ for $N=15$ to $5.6\times 10^{-6}$ for $N=20$, or from $2.6\times 10^{-5}$ for
$N=15$ to $2.6\times 10^{-6}$ for $N=20$ when $m=3$.

Since, by the previous discussion, for large $N$ the level density
is characterized by $\mu$ and $\si$ through the Gaussian law, it is
of interest to compute these parameters in closed form as
functions of $N$ and $m$. In the first place, using the identity
$\tr S_{ij}=m^{N-1}$ and Eqs.~\eqref{Emaxdef} and~\eqref{Emax}, we obtain
\[
\mu=\frac{\tr H}{m^N}=\frac{m+1}{2m}\sum_{i<j}\csc^2(\xi_i-\xi_j)
=\frac{m+1}{12m}N(N^2-1).
\]
Similarly, the formula
\[
\tr(S_{ij}S_{kl})=m^{N-2+2\de_{ik}\de_{jl}+2\de_{il}\de_{jk}}
\]
yields
\begin{align*}
\si^2&=\frac{\tr(H^2)}{m^N}-\frac{(\tr H)^2}{m^{2N}}
=\frac{m^2-1}{4m^2}\sum_{i<j}\csc^4(\xi_i-\xi_j)\\
&=\frac{(m^2-1)N}{8m^2}\sum_{j=1}^{N-1}\csc^4\xi_j\\
&=\frac{m^2-1}{360\ms m^2}\,N(N^2-1)(N^2+11)
\end{align*}
(cf.~Ref.~\cite{CP78}~for the last equality).

The level density is also Gaussian as $N\to\infty$ for the
so-called ``embedded Gaussian ensemble'' (EGOE)~\cite{MF75} in
Random Matrix Theory. Note, however, that in the EGOE this
property is valid provided that the number of one-particle states
tends to infinity faster than $N$. This additional condition
clearly does not hold in our case, since the number of
one-particle states (i.e., $m$) is fixed. Another characteristic
feature of the EGOE is the fact that the nearest-neighbor spacing
distribution $p(s)$ is approximately given by Wigner's law
\[
p(s)=(\pi/2)\ms s\exp(-\pi s^2/4),
\]
as for the classical Gaussian orthogonal ensemble~\cite{Ko01}. On
the other hand, since the HS spin chain is integrable, one would
expect that its nearest-neighbor spacing distribution obey
Poisson's law $p(s)=\e^{-s}$, according to the conjecture of Berry
and Tabor for a generic integrable model~\cite{BT77}. This
conjecture has been verified for a variety of integrable many-body
problems, such as the Heisenberg chain, the $t\ms$-$J$ model, the
Hubbard model~\cite{PZBMM93}, and the chiral Potts
model~\cite{AMV02}. One of the main results of this paper is the
fact that the nearest-neighbor spacing distribution of the HS
chain deviates substantially from both Wigner's and Poisson's
laws.

\begin{figure}[h]
\psfrag{P}[Bc][Bc][1][0]{\begin{footnotesize}$P(s)$\end{footnotesize}}
\psfrag{s}{\begin{footnotesize}$s$\end{footnotesize}}
\includegraphics[width=8cm]{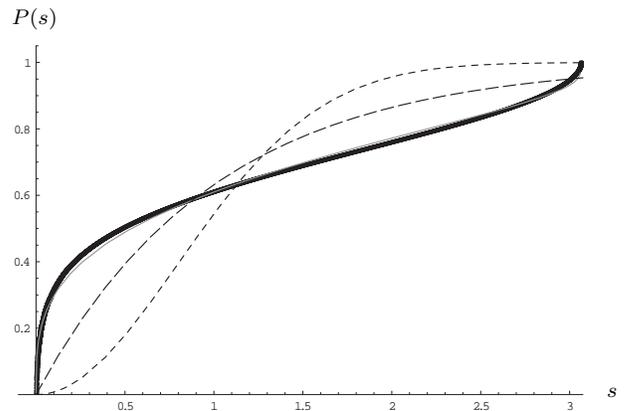}
\caption{Cumulative spacing distribution $P(s)$ and its approximation
$\tP(s)$ (grey line) for $N=26$ and $m=2$. For convenience, we have also represented Poisson's
(long dashes) and Wigner's (short dashes) cumulative distributions.\label{spacings}}
\end{figure}

In order to correctly take into account the effect of the local
level density in the study of $p(s)$, one must first apply to the
``raw'' spectrum the so-called \emph{unfolding}
mapping~\cite{Ha01}. This mapping is defined by decomposing the
cumulative level density $F(E)$ as the sum of a fluctuating part
$F_{\mathrm{f{}l}}(E)$ and a continuous part $\xi(E)$, which is
then used to transform each energy $E_i$, $i=1,\dots,n$, into an
unfolded energy $\xi_i=\xi(E_i)$. The function $p(s)$ is defined
as the density of the normalized spacings
$s_i=(\xi_{i+1}-\xi_i)/\De$, where $\De=(\xi_{n}-\xi_1)/(n-1)$ is
the mean spacing of the unfolded energies. By the previous
discussion, in our case we can take the unfolding mapping $\xi(E)$
as the cumulative Gaussian distribution $G(E)$ with parameters
$\mu$ and $\si$ given by the previous formulas. As for
the level density, to compare the discrete distribution function
$p(s)$ with a continuous distribution it is more convenient to
work with the cumulative spacing distribution $P(s)=\int_0^s
p(x)\ms\diff x$. Our computations for a wide range of values of
$N$ and $m$ show that $P(s)$ is essentially different from either
Poisson's or Wigner's law, since its slope tends to infinity
both as $s\to 0$ and $s\to s_{\mathrm{max}}$, where $s_{\mathrm{max}}$
is the largest spacing. In fact, it turns out that in all cases
$P(s)$ is well approximated by a cumulative distribution of the simple
form
\begin{equation}\label{tP}
\tP(s)=t^\al\big[1-\ga(1-t)^\be\big],
\end{equation}
where $t=s/s_{\mathrm{max}}$ and $0<\al,\be<1$. The parameter $\ga$ is
fixed by requiring that the average spacing be equal to~$1$, with
the result
\begin{equation}\label{ga}
\ga=\Big(\frac 1{s_{\mathrm{max}}}-\frac\al{\al+1}\Big)\Big/B(\al+1,\be+1),
\end{equation}
where $B$ is Euler's Beta function. For instance, for $N=26$ and $m=2$
the largest spacing is $s_{\mathrm{max}}=3.06$, and the best least-squares fit
parameters $\al$ and $\be$ are respectively $0.31$ and $0.23$, with
a mean square error of $4.1\times 10^{-4}$ (see Fig.~\ref{spacings}).

\begin{figure}[h]
\psfrag{N}{\begin{footnotesize}$N$\end{footnotesize}}
\includegraphics[width=8cm]{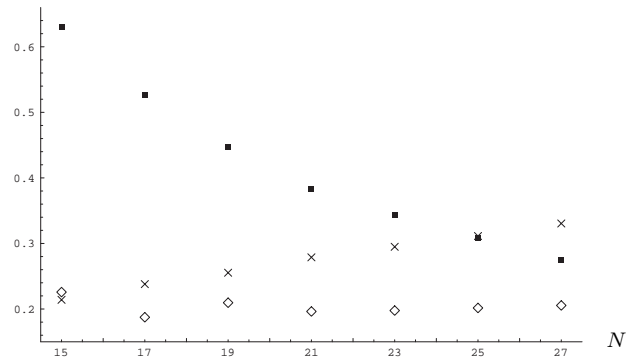}
\caption{Values of $\al$ (box), $\be$ (rhombus), and $s_{\mathrm{max}}/10$ (cross)
for $m=2$ and odd $N$.\label{albesmax}}
\end{figure}

For a fixed value of $m$, the parameters $\al$, $\be$ and $s_{\mathrm{max}}$
vary smoothly with $N\gtrsim 15$, provided that $N$ has a fixed parity\footnote{%
  Our computations show that the number of levels, and hence of different
  spacings, increases monotonically with $N$ of a fixed parity, but decreases
  when $N$ jumps from $2j$ to $2j+1$.}. For instance, in Fig.~\ref{albesmax}
we plot these parameters for $m=2$ and odd $N$ running from $15$ to $27$ (the
plot for even $N$ is very similar). In all cases, the fit of the distribution
\eqref{tP} to the data is quite good, the mean square error never exceeding
$7.4\times 10^{-4}$. We have performed a similar analysis for $m=3$ and
$15\leq N\leq22$, obtaining totally analogous results.

The divergence of the nearest-neighbor spacing distribution $p(s)$
for small $s$ is probably related to the flatness of the tail of the
Gaussian distribution. It could also be argued that, since
the Haldane--Shastry chain is completely integrable,
the full spectrum is a superposition of the spectra of the
Hamiltonian restricted to subspaces of common eigenfunctions of a
suitable family of commuting first integrals. It is well known, in this respect,
that a superposition of a large number of unrelated spectra leads to a
sharp increase in the number of very small spacings~\cite{RP60}.
On the other hand, we do not have a clear explanation of the fact that
$p(s)$ also diverges when $s$ approaches the largest spacing $s_{\mathrm{max}}$.
This fact, which certainly deserves further study, could be a characteristic
property of all spin chains of Haldane--Shastry type.

\begin{figure}[t]
\psfrag{P}[Bc][Bc][1][0]{\begin{footnotesize}$P(s)$\end{footnotesize}}
\psfrag{s}{\begin{footnotesize}$s$\end{footnotesize}}
\includegraphics[width=8cm]{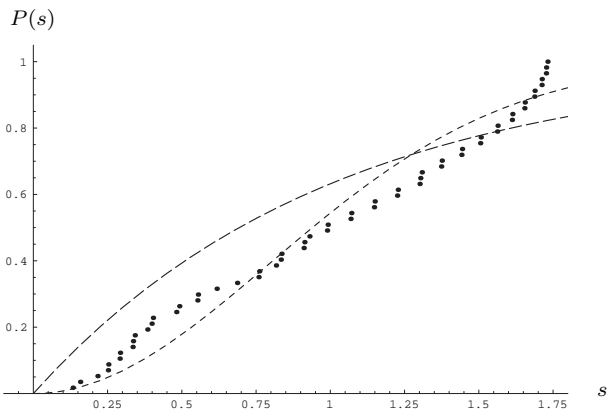}
\caption{Cumulative spacing distribution $P(s)$ (solid dots) for states with
  zero total spin and odd parity when $N=13$ and $m=2$. For comparison
  purposes, we have represented Poisson's (long dashes) and Wigner's (short
  dashes) cumulative distributions.%
\label{fixspin}}
\end{figure}

Our results also imply that Berry and Tabor's conjecture does not
hold for the HS spin chain, even if we restrict ourselves to a
subspace of the whole Hilbert space with well-defined quantum
numbers. Indeed, the nearest-neighbor spacing distribution of the
superposition of even a small number of spectra with
Poisson-distributed spacings must also be of Poisson
type~\cite{RP60}. As an illustration of these assertions, we
present in Fig.~\ref{fixspin} a plot of the cumulative spacing
distribution corresponding to the restriction of the
Hamiltonian~\eqref{HS} to the subspace with zero total spin and
odd parity for $N=13$ and $m=2$, obtained by a numerical
computation of the spectrum of $H$ restricted to this subspace. It
is apparent from this plot that $P(s)$ is neither Poissonian nor
of Wigner type, and that it is well approximated by a function of
the form~\eqref{tP} for spacings $s\gtrsim0.25$. It is also
clearly noticeable that $p(s)$ tends to infinity as $s$ approaches
the maximum spacing $s_{\mathrm{max}}\simeq1.73$.

The non-Poissonian behavior of the spacing distribution could in principle be
due to finite-size effects \cite{KD05}. Although this possibility should be
explored in more detail, our data clearly show that the cumulative spacing
distribution $P(s)$ is of the form~\eqref{tP} for a wide range of values of
$N\leq27$.

Note, finally, that an interesting integrable model not obeying the
Berry--Tabor conjecture has been recently constructed in Ref.~\cite{RDGR04}.
In contrast with the HS spin chain, the latter model is a non-generic element
of a class depending on a large number of parameters, and involves many-body
interactions.
%

\medskip
\begin{acknowledgments}
This work was partially supported by the DGI under grant No.~BFM2002--02646.
The authors would like to thank J.~Retamosa for several helpful discussions.
\vspace*{.5cm}
\end{acknowledgments}

\end{document}